\documentclass[11pt]{article}
\usepackage{moriond_warburton,epsfig}

\def\Journal#1#2#3#4{{#1} {\bf #2}, #3 (#4)}

\def\PRL{\em Phys. Rev. Lett.}
\def\PRD{{\em Phys. Rev.} D}

\def\be{\begin{equation}}
\def\ee{\end{equation}}
\def\bea{\begin{eqnarray}}
\def\eea{\end{eqnarray}}

\begin{document}

\vspace*{0.5cm}
\begin{flushright}
CLNS~03/1830\\
MCGILL~03/11
\end{flushright}

\vspace*{4cm}
\title{$|V_{cb}|$ AND $|V_{ub}|$ FROM CLEO}

\author{{\sc Andreas Warburton}\\
        (representing the CLEO Collaboration)\\
\vspace{2cm}}
\address{Laboratory for Elementary-Particle Physics\\
Cornell University\\
Ithaca, New York\ \ 14853\\
USA}

\address{Department of Physics\\
McGill University\\
Montr\'{e}al, Qu\'{e}bec\ \ H3A 2T8\\
Canada}

\maketitle\abstracts{ I report on studies of inclusive and exclusive
  semileptonic $b \to c\ell\nu$ and $b \to u\ell\nu$ decays in 9.7
  million $B\bar B$ events accumulated with the CLEO detector in
  symmetric $e^+e^-$ collisions produced in the Cornell Electron
  Storage Ring (CESR).  Various experimental techniques, including the
  study of spectral moments and the inference of neutrino candidates
  by exploiting the hermeticity of the CLEO detector, are used in
  conjunction with theoretical calculations to provide estimates of
  the CKM matrix elements $|V_{cb}|$ and $|V_{ub}|$.  }

\vspace*{1.0in}
\begin{center}
{\it Invited talk presented at the 2003 QCD and Hadronic
Interactions \\ session of the XXXVIII$^{th}$ Rencontres de
Moriond, \\ Les Arcs 1800, Savoie, France}
\end{center}

\newpage

\section{Introduction}

High-precision measurements to help overconstrain the
Cabibbo-Kobayashi-Maskawa\cite{ckm} (CKM) quark-mixing matrix through
its elements $|V_{cb}|$ and $|V_{ub}|$ are categorically blighted by
the effects of long-distance QCD.  Arguably, the impetus behind many
recent theoretical advances in the non-perturbative QCD physics of
meson decay has been to enable access to fundamental electroweak
quantities like $|V_{cb}|$ and $|V_{ub}|$.  The $4\pi$ solenoidal CLEO
detector\cite{cleo}, which comprises tracking chambers, a CsI
electromagnetic calorimeter, and muon systems, has enabled several new
measurements\cite{bsg,hadmom,inclvub,exclvcb,lepmom,exclvub} of
$|V_{cb}|$ and $|V_{ub}|$ through both inclusive and exclusive studies
of moments and rates observed in semileptonic decays of $B$ mesons
produced in $e^+e^-$ collisions at the Cornell Electron Storage Ring
(CESR).  Due to a dearth of space, we outline here briefly a new
determination of $|V_{ub}|$ from extensive studies of exclusive $B \to
[\pi,\rho,\omega,\eta]\ell\nu$ decays\cite{exclvub}, and refer the
reader to recent publications\cite{bsg,hadmom,inclvub,exclvcb,lepmom}
for details of the other CLEO CKM measurements presented.

\section{Event Reconstruction and Selection}

Exclusive semileptonic decay studies are made difficult by the
non-interacting neutrino, the kinematics of which can be inferred
using the hermeticity of the CLEO detector by reconstructing missing
energy ($E_{\rm{miss}}\equiv 2E_{\rm{beam}}-\sum E_i$) and missing
momentum ($\vec{P}_{\rm{miss}}\equiv -\sum\vec{p}_i$) in every
event\cite{exclvub}.  Within resolution, the signal neutrino combined
with its companion charged lepton ($\ell$) and meson ($m$) should
satisfy constraints on energy, $\Delta E \equiv
(E_\nu+E_\ell+E_m)-E_{\rm{beam}} \approx 0$, and on momentum,
$M_{m\ell\nu} \equiv [E_{\rm{beam}}^2 -
|\alpha\vec{p}_\nu+\vec{p}_\ell+\vec{p}_m|^2]^\frac{1}{2} \approx
M_B$, where $\alpha$ is chosen to force $\Delta E=0$.  Events are
examined with total charge $|\Delta Q| =$~0~or~1, and we reconstruct
the momentum transfer $q^2=M_{W^*}^2=(p_\nu+p_\ell)^2$ from the
missing momentum and the charged lepton's kinematics.  Candidate
charged leptons are required to have momenta $p_\ell
>$~1.0$\,$(1.5)~GeV/$c$ for the pseudoscalar (vector) reconstructions.

\section{Extraction of Branching Fractions}

\begin{figure}
\begin{center}
\epsfig{figure=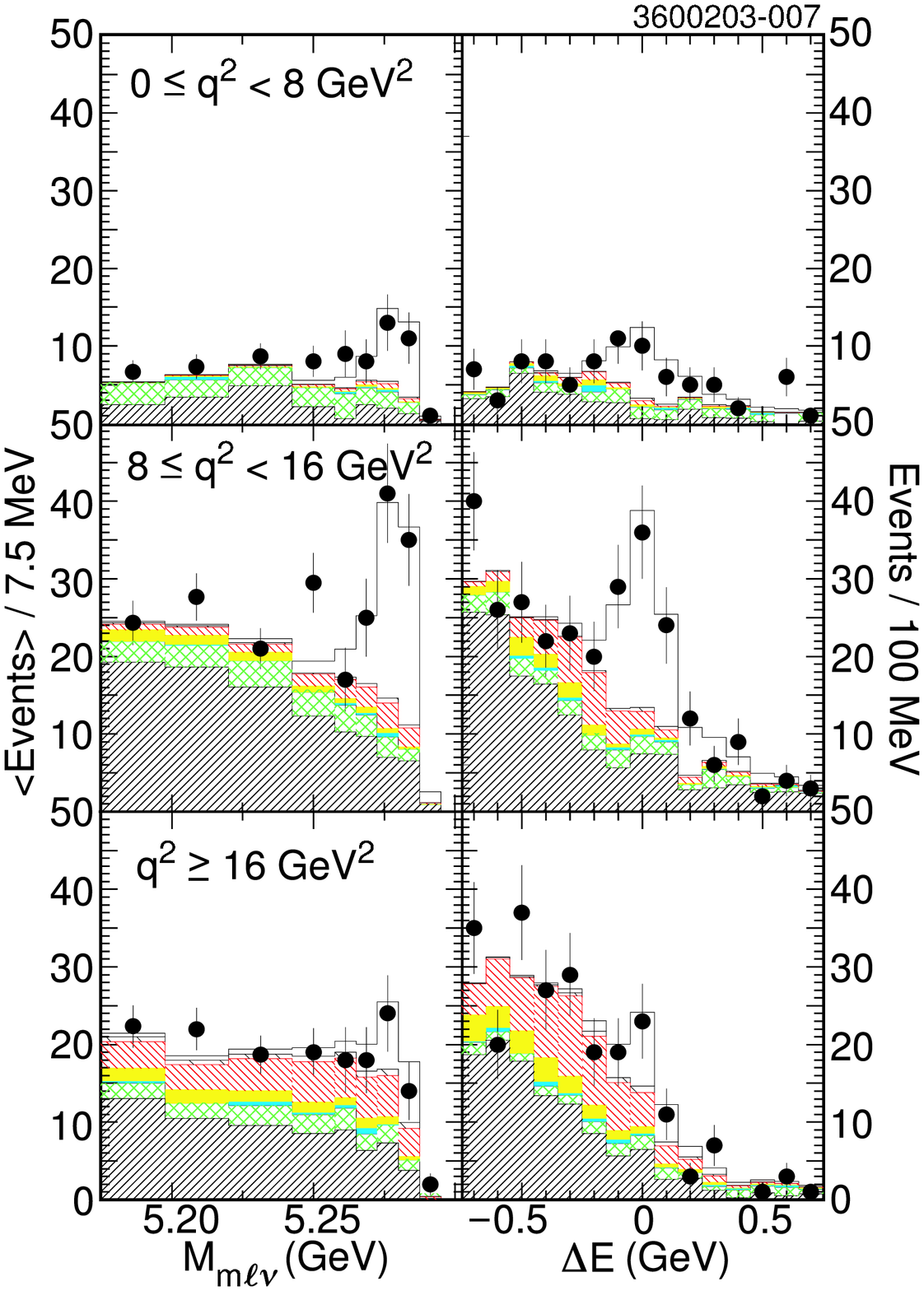,height=4.5in}
\hskip 0.3in
\epsfig{figure=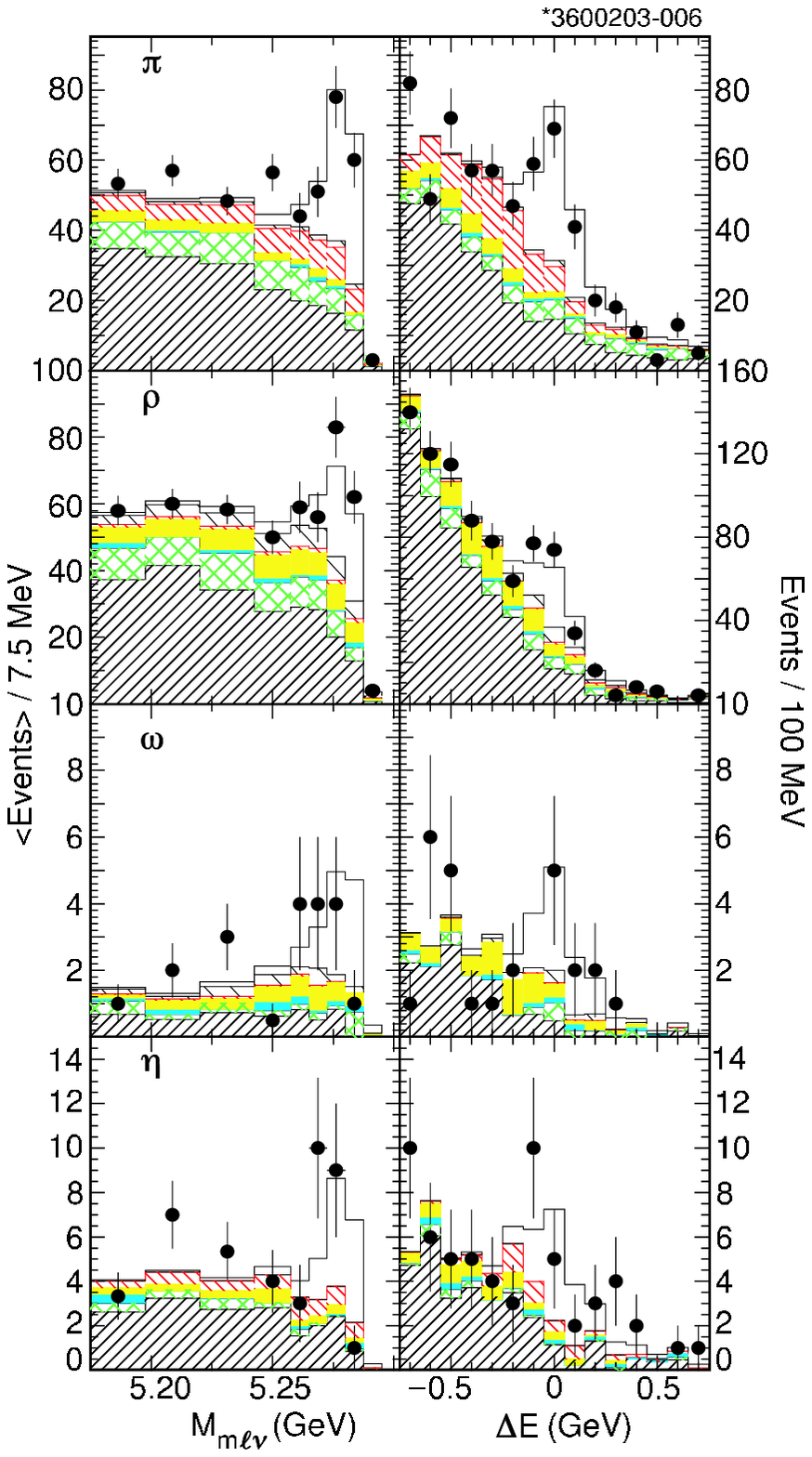,height=4.5in}
\end{center}
\caption{[left] $M_{m\ell\nu}$ and $\Delta E$ observables in
  the $\Delta E$ and $M_{m\ell\nu}$ signal band for the three $q^2$
  bins (rows), requiring $\Delta Q=0$ for the combined $\pi^\pm,
  \pi^0$ modes; [right] the same observables summed over the entire
  $q^2$ range for the combined $\pi$ modes (top), $\rho$ modes (row
  2), $\omega$ (row 3), and $\eta$ (bottom) modes.  The points are the
  on-resonance data.  The histogram components, from bottom to top,
  are $b\to c$ (fine $45^\circ$ hatch); continuum (grey or green cross
  hatch); fake leptons (cyan or dark grey); feed down from other $B\to
  X_u\ell\nu$ modes (yellow or light grey); for the $\pi^\pm$ and
  $\pi^0$ modes, cross feed from the vector and $\eta$ modes into the
  $\pi$ modes (red or black fine $135^\circ$ hatch), cross feed among
  the $\pi$ modes (coarse $135^\circ$ hatch); for the $\rho^\pm$ and
  $\rho^0$ modes, cross feed from the $\pi$ and $\eta$ modes into the
  $\rho$ modes (red or black fine $135^\circ$ hatch), cross feed among
  the vector modes (coarse $135^\circ$ hatch); for the $\eta$ mode,
  there is only a single cross-feed component from the non-$\eta$
  modes (red or black fine $135^\circ$ hatch); and signal (open).  The
  normalizations are those from the nominal fit.}
\label{fig:bmmasses}
\end{figure}

We performed a maximum likelihood fit\cite{exclvub} in bins of the
observables $\Delta E$, $M_{m\ell\nu}$, $\Delta Q$, $2\pi\,(3\pi)$
meson mass ranges in the $\rho\ell\nu\,(\omega\ell\nu)$ modes, and
$q^2$; in total, the nominal fit involved 7 signal-mode topologies,
comprised 259 bins, and had a $\chi^2$ probability of 0.48.
Projections of the nominal fit in the $\Delta E$ and $M_{m\ell\nu}$
variables in bins of $q^2$ and decay mode, as well as the various
signal and background fit components, are indicated in
Fig.~\ref{fig:bmmasses}.  The experimental branching-fraction results
are summarized in Tab.~\ref{tab:nomfit_br}.  By examining decay rates
differentially in $q^2$, we achieve a marked reduction in the
sensitivity to theoretical model uncertainties originating from the
form-factor shapes needed to simulate the signal and cross-feed fit
components\cite{exclvub}.  We emphasize that theoretical form factors
do {\em not} constrain the relative rates across $q^2$ bins.  For the
$B^0\to\pi^-\ell^+\nu$ ($B^0\to\rho^-\ell^+\nu$) mode, we append to
the total branching fractions listed in Tab.~\ref{tab:nomfit_br}
residual theoretical form-factor uncertainties of $[\pm 0.01 \pm
0.07]\times 10^{-4}$ ($[\pm 0.41 \pm 0.01]\times 10^{-4}$) due to the
signal and cross-feed components, respectively.  To the
$B^+\to\eta\ell^+\nu$ total branching fraction, we append a model
dependence\cite{isgw2} uncertainty of $0.09\times 10^{-4}$.

\begin{table}
\caption{Summary of branching fractions from the nominal fit using the
Ball'01\protect\cite{ball01}, Ball'98\protect\cite{ball98},
and ISGW$\,$II\protect\cite{isgw2} form factors for the $\pi$, $\rho$,
and $\eta$ modes, respectively.  The first uncertainties are statistical
and the second systematic; the theoretical form-factor uncertainties are not
listed\protect\cite{exclvub}.  The results for the fits with more
restrictive charged-lepton momentum criteria in the vector modes are
also shown.  The $q^2$ intervals are specified in units of GeV$^2$.}
\begin{center}
\begin{tabular}{ccccc}  \hline\hline
Mode & ${\cal B}_{q^2\,\mathrm{interval}}$
     &\multicolumn{3}{c}{Vector-Mode Charged-Lepton Momentum} \\ 
& $\times 10^4$ & $p_\ell>1.5\ \mathrm{GeV}/c$ 
& $p_\ell>1.75\ \mathrm{GeV}/c$  & $p_\ell>2.0 \ \mathrm{GeV}/c$ \\
\hline
$B^0\to\pi^-\ell^+\nu$  & ${\cal B}_{\rm{total}}$ & 
                          $1.33\pm 0.18\pm 0.11$ &
                          $1.31\pm 0.18\pm 0.11$ &
                          $1.32\pm 0.18\pm 0.12$ \\
                        & ${\cal B}_{< 8}$ &
                          $0.43\pm 0.11\pm 0.05$ &
                          $0.43\pm 0.11\pm 0.05$ &
                          $0.42\pm 0.11\pm 0.05$ \\
                        & ${\cal B}_{8-16}$ &
                          $0.65\pm 0.11\pm 0.07$ &
                          $0.65\pm 0.11\pm 0.07$ &
                          $0.66\pm 0.11\pm 0.07$ \\
                        & ${\cal B}_{\ge16}$ &
                          $0.25\pm 0.09\pm 0.04$ &
                          $0.24\pm 0.09\pm 0.04$ &
                          $0.24\pm 0.09\pm 0.05$ \\ \hline
$B^0\to\rho^-\ell^+\nu$ & ${\cal B}_{\rm{total}}$ &
                          $2.17\pm 0.34\;^{+0.47}_{-0.54}$ &
                          $2.34\pm 0.34\;^{+0.43}_{-0.51}$ &
                          $2.29\pm 0.35\;^{+0.40}_{-0.49}$\\
                        & ${\cal B}_{< 8}$ &
                          $0.43\pm 0.20\;^{+0.23}_{-0.23}$ &
                          $0.50\pm 0.20\;^{+0.21}_{-0.22}$ &
                          $0.62\pm 0.22\;^{+0.22}_{-0.23}$\\
                        & ${\cal B}_{8-16}$ &
                          $1.24\pm 0.26\;^{+0.27}_{-0.33}$ &
                          $1.32\pm 0.26\;^{+0.26}_{-0.29}$ &
                          $1.11\pm 0.25\;^{+0.23}_{-0.25}$\\
                        & ${\cal B}_{\ge16}$ &
                          $0.50\pm 0.10\;^{+0.08}_{-0.11}$ &
                          $0.52\pm 0.10\;^{+0.08}_{-0.10}$ &
                          $0.56\pm 0.10\;^{+0.07}_{-0.09}$\\ \hline
$B^+\to\eta\ell^+\nu$   & ${\cal B}_{\rm{total}}$ &
                          $0.84\pm 0.31\pm 0.16$  &
                          $0.84\pm 0.31\pm 0.16$  &
                          $0.83\pm 0.31\pm 0.15$ \\ \hline\hline
\end{tabular}
\end{center}
\label{tab:nomfit_br}
\end{table}

\section{Extraction of $|V_{ub}|$}

\begin{figure}
\begin{center}
\epsfig{figure=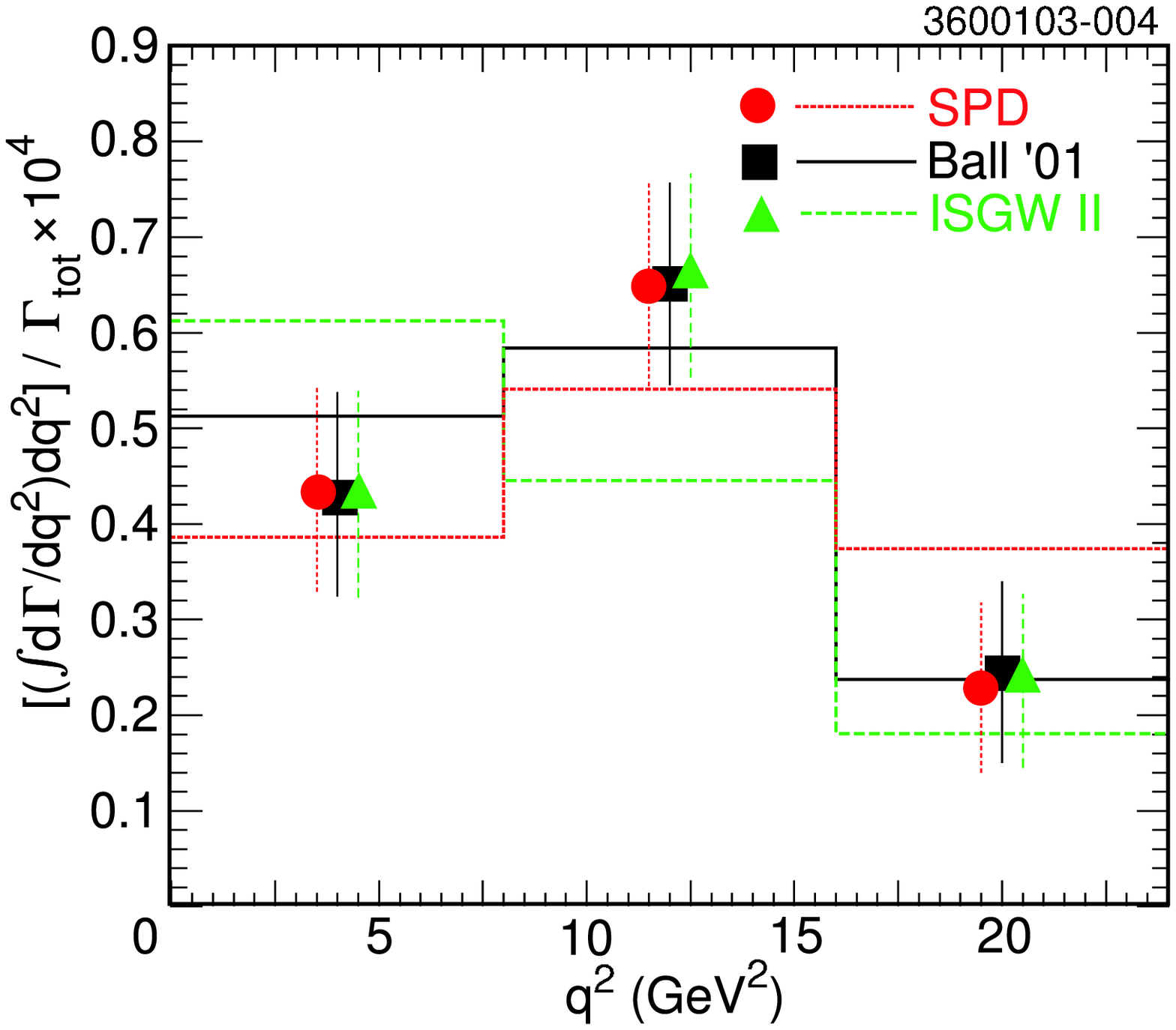,height=2.2in}
\hskip 0.3in
\epsfig{figure=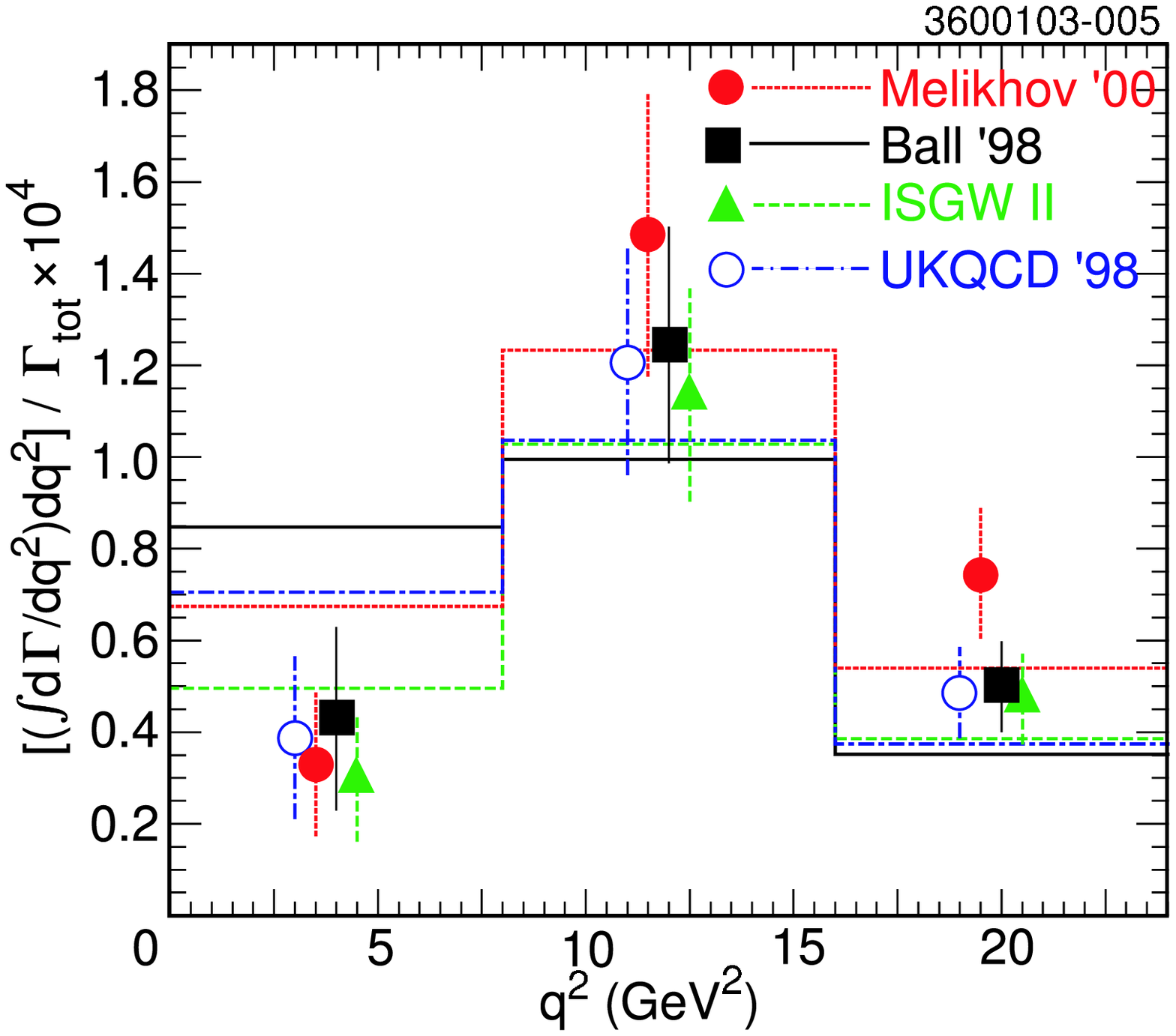,height=2.2in}
\end{center}
\caption{Measured branching fractions (points) in the restricted $q^2$
  intervals for $B^0\to\pi^-\ell^+\nu$ [left] and
  $B^0\to\rho^-\ell^+\nu$ [right], and the best fit to the predicted
  $d\Gamma/dq^2$ (histograms) for various models (SPD -- skewed parton
  distributions\protect\cite{spd}; Melikhov'00 -- a relativistic quark
  model\protect\cite{meli00}; UKQCD'98 -- a lattice QCD
  calculation\protect\cite{ukqcd98}) used to extract both rates and
  $|V_{ub}|$.  The small horizontal offsets in the data points have
  been introduced for clarity.}
\label{fig:Vub_fits}
\end{figure}

We extract values of $|V_{ub}|$ using several approaches\cite{exclvub}
involving the measured rates for $\pi\ell\nu$ only, those for
$\rho\ell\nu$ only, and an optimized combination of both mode types.
Results from fits over the totality of phase space are indicated in
Fig.~\ref{fig:Vub_fits}, where the quality of fit can aid in
discriminating between form-factor models by measuring how well the
form-factor shapes describe the data.  For example, in the
$\pi\ell\nu$ case, the ISGW$\,$II\cite{isgw2} model receives only a
$\chi^2$ probability in the range $1-3$\% in our various fits,
suggesting that this model is likely to provide less reliable
$|V_{ub}|$ extractions.
% in the $\pi\ell\nu$ mode.

As a preferred alternative, we employ a method to determine $|V_{ub}|$
that reduces modeling assumptions by exploiting theoretical QCD
calculations solely in their $q^2$ regions of validity.  In each of
the $\pi\ell\nu$ and $\rho\ell\nu$ cases, we use form-factor shape and
normalization results from light-cone sum rules (LCSR) QCD
calculations\cite{exclvub} in the region $q^2 < 16$~GeV$^2$ and rate
calculations from lattice QCD (LQCD) studies\cite{exclvub} in the
complementary $q^2 \geq 16$~GeV$^2$ bin.  In each of the $\pi\ell\nu$
and $\rho\ell\nu$ modes, we average the LCSR and LQCD results, taking
into account correlated systematic uncertainties; we then combine the
$|V_{ub}|$ results from the two mode types using an optimized
weighting\cite{exclvub}.  We find
%\begin{equation}
$|V_{ub}| = \left[ 3.17 \pm 0.17 \;^{+0.16}_{-0.17} \; ^{+0.53}_{-0.39}
\pm 0.03 \right] \times 10^{-3}$,
%\end{equation}
where the uncertainties are statistical, experimental systematic,
theoretical systematic based on the LCSR and LQCD uncertainties, and
the $\rho\ell\nu$ form-factor shape uncertainty, respectively.  Our
results minimize our reliance on modeling and are consistent with
previous rate and $|V_{ub}|$ measurements.  Significant progress in
the extraction of $|V_{ub}|$ from exclusive decays will require major
theoretical improvements.

\section*{Acknowledgments}
My colleagues in the CLEO collaboration and the CESR staff made these
results possible.  I am also grateful to McGill University for
financial support.
% and to the conference organizers
% for providing a stimulating meeting environment.

\end{document}